\RequirePackage{lineno}
\documentclass[aps,preprint,showpacs,amsmath,amssymb,epsfig]{revtex4}
\usepackage{graphicx}% Include figure files
\usepackage[usenames]{color}
\usepackage{epstopdf}
\usepackage{bm}% bold math

\usepackage[colorlinks=true, urlcolor=navyblue, linkcolor=navyblue, citecolor=navyblue]{hyperref}
\definecolor{navyblue}{rgb}{0,0.08,0.45}
\begin{document}

\preprint{SLAC--PUB--16959}
\preprint{JLAB-PHY-17-2394}

\title{Implications of the Principle of Maximum Conformality for  the QCD Strong Coupling}

\author{Alexandre Deur$^{1}$}
\email[email:]{deurpam@jlab.org}

\author{ Jian-Ming Shen$^{2}$}
\email[email:]{cqusjm@cqu.edu.cn}

\author{Xing-Gang Wu$^{2}$}
\email[email:]{wuxg@cqu.edu.cn}

\author{Stanley J. Brodsky$^{3}$}
\email[email:]{sjbth@slac.stanford.edu}

\author{Guy F. de T\'eramond$^{4}$}
\email[email:]{gdt@asterix.crnet.cr}

\address{$^1$Thomas Jefferson National Accelerator Facility, Newport News, VA 23606}
\address{$^2$Department of Physics, Chongqing University, Chongqing 401331, P.R. China}
\address{$^3$SLAC National Accelerator Laboratory, Stanford University, Stanford, California 94309, USA}
\address{$^4$Universidad de Costa Rica, 11501 San Jos\'e, Costa Rica}

%\date{\today}

\begin{abstract}

The Principle of Maximum Conformality (PMC) provides scale-fixed perturbative QCD predictions which are independent of the choice of the renormalization scheme, as well as the choice of the initial renormalization scale. In this article, we will test the PMC by comparing its predictions for the strong coupling $\alpha^s_{g_1}(Q)$,  defined from the Bjorken sum rule, with predictions using conventional pQCD scale-setting. The two results are found to be compatible with each other and with the available experimental data.  However, the PMC provides a significantly more precise determination, although its domain of applicability ($Q \gtrsim 1.5$ GeV) does not extend to as small values of momentum transfer  as that of a conventional pQCD analysis ($Q \gtrsim 1$ GeV).  We suggest that the PMC range of applicability could  be improved by a modified intermediate scheme choice or using a single effective PMC scale.

\end{abstract}

\pacs{12.38.Aw, 12.38.Lg}

\maketitle

\section{introduction}

The gauge theory of the strong interactions, Quantum Chromodynamics (QCD) is defined to provide objective predictions for physical observables; its predictions should not depend on arbitrary theory conventions, such as the choice of the gauge or the choice of renormalization scheme (RS).  However, conventional calculations are typically carried out using a perturbative formalism where the truncated high-order predictions are RS-dependent.  Furthermore, the $n! $ growth of the $n^{th}$ order coefficient of the resulting series --the renormalon problem~\cite{Zakharov:1997xs}-- makes the convergence of the series problematic, even at high momentum transfer where the QCD coupling $\alpha_s$ becomes small.  A methodology to solve these problems has been developed, starting with the BLM procedure~\cite{Brodsky:1982gc}, extended by  Commensurate Scale Relations~\cite{Brodsky:1994eh}, and culminating with the Principle of Maximum Conformality (PMC)~\cite{Brodsky:2011ta, Brodsky:2012rj, Brodsky:2011ig, Mojaza:2012mf, Brodsky:2013vpa}.

The PMC provides a systematic method to eliminate the renormalization scheme and scale dependences of conventional pQCD predictions for high-momentum transfer processes. It reduces in the Abelian limit ($N_c\to 0$)~\cite{Brodsky:1997jk}
to the QED Gell-Mann-Low scale-setting method~\cite{GellMann:1954fq}, and it provides the underlying principle for the BLM procedure, extending it unambiguously to all orders consistent with  renormalization group methods.  The PMC has a solid theoretical foundation, satisfying renormalization group invariance~\cite{Wu:2013ei, Wu:2014iba} and all  other self-consistency conditions, such as reflexivity, symmetry, and transitivity derived from the renormalization group~\cite{Brodsky:2012ms}.

The PMC scales in the pQCD series are determined  by shifting the arguments of the strong coupling $\alpha_s(Q^2) $ at each order $n$ to eliminate all occurrences of the non-conformal $\{\beta_i\}$-terms.  The  terms involving $\{\beta_i\}$  are identified at each order using the  recursive pattern dictated by the renormalization group equation (RGE)~\cite{Mojaza:2012mf, Brodsky:2013vpa}. This unambiguous procedure determines the scales $Q_n$ of the strong coupling at each specific order.  As in QED, the PMC scales have a physical meaning in the sense that they are proportional to the virtuality of the gluon propagators at each given order, as well as setting the effective number $n_f$ of active quark flavors.  After applying the PMC, the divergent renormalon series disappear, and the pQCD convergence is automatically improved.  After normalizing the coupling to experiment at a single scale, the PMC predictions become scheme-independent. The PMC has been successfully applied to many high-energy processes; see, e.g.,  Ref.~\cite{Wang:2017kyd}.

In this paper, we shall test the applicability of the PMC by comparing its prediction for the evolution of the QCD strong coupling $\alpha_s(Q)$ to the corresponding prediction
based on conventional scale-setting, where the renormalization scale at each order is estimated as a typical momentum transfer of the process and where arbitrary range and systematic error are assigned to estimate the uncertainty of the fixed-order pQCD predictions.

The PMC  will be applied in this paper in order to determine the behavior of the running coupling $\alpha_{g_1}(Q)$, using  the $\overline{\rm MS}$-scheme as an auxiliary RS.   The coupling $\alpha_{g_1}(Q)$  is an ``effective charge"~\cite{Grunberg:1980ja} -- i.e., an observable -- defined from the Bjorken sum rule~\cite{Bjorken:1966jh, Bjorken:1969mm}.  It involves the spin-dependent $g_1$ structure function;  hence, its name. The PMC prediction for $\alpha_{g_1}(Q)$ is RS-independent, whereas the conventional pQCD calculation  of $\alpha_{g_1}(Q)$ retains RS-dependence,  typically chosen as the $\overline{\rm MS}$ scheme.

This article is organized as follow: In Sec.~\ref{alpha_s in MSbar}, we recall the formalism which defines the $\alpha_{\overline{\rm MS}}(Q)$ renormalization scheme and the pQCD expansion for the effective charge  $\alpha_{g_1}(Q)$ using conventional pQCD scale-setting.  In Sec.~\ref{alpha_s from PMC}, we provide the formulae which allow the computation of $\alpha_{g_1}(Q)$ using the PMC.  In Sec.~\ref{PMC test}, we compare the two calculations.  In Sec.~\ref{matching}, we discuss the possibility of using the PMC in a procedure that employs $\alpha_{s}$ to relate the fundamental QCD parameter $\Lambda_{\overline{\rm MS}}$ to hadron masses or, equivalently, to the confinement scale $\kappa$ emerging from the Light-Front Holographic QCD approach to nonperturbative QCD~\cite{Brodsky:2014yha}. We summarize the results in the final  section.

\section{ PQCD computation of the effective charge $\alpha_{g_1}$ in the $\overline{\rm MS}$ scheme\label{alpha_s in MSbar}}

In the $\overline{\rm MS}$-scheme, the effective charge $\alpha_{g_1}$ has the leading-twist perturbative expansion \cite{Deur:2005cf}:
\begin{equation}
\label{eq:msbar to g_1}
\frac{\alpha_{g_{1}}(Q)}{\pi}=\sum_{i\geq 1} a_i \left(\frac{\alpha_{\overline{\rm MS}}(Q)}{\pi}\right)^i.
\end{equation}
The perturbative coefficients $a_i$ are known up to four loops~\cite{Baikov:2010je, Baikov:2012zm}. (The values are given explicitly in Section~\ref{alpha_s from PMC},  Eq.~\ref{convseries}.) The definition of $\alpha_{g_1}$ stems from the Bjorken sum rule~\cite{Bjorken:1966jh, Bjorken:1969mm}.
At leading-twist:
 \begin{align}
\int_0^{1^-} g_1^{p-n}(x_{Bj},Q) \ dx_{Bj} &=
\frac{g_a}{6}\left[1- \sum_{i\geq 1} a_i \left(\frac{\alpha_{\overline{\rm MS}}(Q)}{\pi}\right)^i \right]  
\equiv \frac{g_a}{6}\left[1-\frac{\alpha_{g_1}(Q)}{\pi} \right] {\color{blue},}
\label{eqn:bj}
\end{align}
where the integration runs over the Bjorken scaling variable $x_{Bj}$. 
The nucleon axial charge is $g_a$ and  the label {\small \emph{p-n}} indicates the isovector part 
of the spin structure function $g_1$. The Bjorken integral is well measured, including  the transition 
region between perturbative to nonperturbative QCD~\cite{Deur:2004ti}. The $Q^2$-evolution of 
the strong coupling $\alpha_{\overline{\rm MS}}(Q^2)$ in the $\overline{\rm MS}$-scheme is governed by the RGE:
\begin{equation}
\label{beta MSbar}
Q^{2}\frac{\partial}{\partial Q^{2}}\left(\frac{\alpha_{s}}{4\pi}\right) =\beta\left(\alpha_{s}\right)=-\sum_{n\geq0} \left(\frac{\alpha_{s}}{4\pi}\right)^{n+2}\beta_{n},
\end{equation}
which is known up to 5-loops:
\begin{eqnarray}
\beta_{0} &=& 11-\frac{2}{3}n_{f}, \nonumber \\
\beta_{1} &=&102-\frac{38}{3}n_{f}, \nonumber \\
\beta_{2} &=& \frac{2857}{2}-\frac{5033}{18}n_{f}+\frac{325}{54}n_{f}^{2}, \nonumber \\
\beta_{3} &=& \frac{149753}{6}+3564 {\xi_3} -\bigg(\frac{1078361}{162}+\frac{6508}{27}\xi_3 \bigg)n_{f}
+\bigg(\frac{50065}{162}+\frac{6472}{81} \, \xi_3 \bigg)n_{f}^{2}+\frac{1093}{729}n_{f}^{3}, \nonumber \\
\beta_4 &=& \frac{8157455}{16} + \frac{621885}{2}\xi_3 - \frac{88209}{2}\xi_4 -288090 \xi_5
+\bigg(-\frac{336460813}{1944} -\frac{4811164}{81}\xi_3 + \frac{33935}{6}\xi_4  \nonumber \\
&& +\frac{1358995}{27}\xi_5 \bigg) n_f + \bigg(\frac{25960913}{1944} + \frac{698531}{81}\xi_3
-\frac{10526}{9}\xi_4 - \frac{381760}{81}\xi_5 \bigg)n_f^2+   \nonumber \\
&& \bigg(-\frac{630559}{5832} -\frac{48722}{243}\xi_3 + \frac{1618}{27}\xi_4 + \frac{460}{9}\xi_5 \bigg)n_f^3
+\bigg(\frac{1205}{2916} -\frac{152}{81}\xi_3 \bigg)n_f^4,  \nonumber
\end{eqnarray}
where $\xi_n$ is the Riemann zeta function \cite{Baikov:2016tgj, Luthe:2016ima}. The coefficients $\beta_i$ are expressed utilizing the $\overline{\rm MS}$-scheme except for $\beta_0$ and $\beta_1$ which are scheme independent.

Solving Eq.~(\ref{beta MSbar}) iteratively yields the approximate five-loop expression of $\alpha_{\overline{\rm MS}}^{\rm pQCD}$~\cite{Kniehl:2006bg},
\begin{eqnarray}
\alpha^{\rm pQCD}_{\overline{\rm MS}}(Q) &=& \frac{4\pi}{\beta_{0}t} \biggl[1-\frac{\beta_{1}}{\beta_{0}^{2}}\frac{\mbox{ln}(t)}{t} +\frac{\beta_{1}^{2}}{\beta_{0}^{4}t^2}\bigg(\mbox{ln}^2(t)-\mbox{ln}(t)
 -1+\frac{\beta_{2}\beta_{0}}{\beta_{1}^{2}}\bigg)+ \frac{\beta_{1}^{3}}{\beta_{0}^{6}t^{3}}\biggl(-\mbox{ln}^3(t)+\frac{5}{2}\mbox{ln}^2(t)  \nonumber\\
&& +2 \mbox{ln}(t)-\frac{1}{2}-3\frac{\beta_{2}\beta_{0}}{\beta_{1}^{2}}\mbox{ln}(t)+\frac{\beta_{3}\beta_{0}^{2}}{2\beta_{1}^{3}}\biggr)
+ \frac {\beta_1^4} {\beta_0^8 t^4} \bigg(\mbox{ln}^4(t) -\frac{13} {3} \mbox{ln}^3(t) -\frac{3} {2}  \mbox{ln}^2(t)  +4 \mbox{ln}(t) \nonumber\\
&& +\frac{7}{6} +  \frac {3\beta_2 \beta_0}{\beta_1^2} \left( 2 \, \mbox{ln}^2(t) - \mbox{ln}(t) -1\right)  -  \frac{\beta_3\beta_0^2}{\beta_1^3}\left(2 \, \mbox{ln}(t) + \frac{1}{6}\right) +\frac{5\beta_2^2 \beta_0^2}{3\beta_1^4}+\frac{\beta_4 \beta_0^3}{3 \beta_0^4} \biggr) \biggr]   \nonumber\\
&& +\cdots
\label{eq:alpha_s},
\end{eqnarray}
where $t=\mbox{ln}\left(Q^{2}/\Lambda_s ^{2}\right)$ and $\Lambda_s$ is the asymptotic scale. Eqs.~(\ref{eq:msbar to g_1}) to~(\ref{eq:alpha_s}) allow us to compute $\alpha_{g_1}(Q)$ in the pQCD domain. Although $\alpha_{g_1}$ is an observable, the $\overline{\rm MS}$ RS-dependence remains in its pQCD approximant due to the truncations of Eqs.~(\ref{eq:msbar to g_1}) to~(\ref{eq:alpha_s}).

\section{PMC scale-setting for $\alpha_{g_1}(Q^2)$  \label{alpha_s from PMC}}

Following the basic PMC procedure, we first identify the conformal and nonconformal pQCD contributions  for $\alpha_{g_1}$. The corresponding expression (\ref{eq:msbar to g_1}) is then reorganized as~\cite{Brodsky:2013vpa, Shen:2016dnq}
\begin{eqnarray}
\frac{\alpha_{g_1}(Q)}{\pi} &=& r_{1,0} \frac{\alpha_{\overline{\rm MS}}(Q)}{\pi} + (r_{2,0} + \beta_0 r_{2,1}) \left(\frac{\alpha_{\overline{\rm MS}}(Q)}{\pi}\right)^2 +(r_{3,0} + \beta_1 r_{2,1} + 2\beta_0 r_{3,1} +  \nonumber\\
&& \beta_0^2 r_{3,2}) \left(\frac{\alpha_{\overline{\rm MS}}(Q)}{\pi}\right)^3 +(r_{4,0} + \beta_2 r_{2,1} + 2\beta_1 r_{3,1} + \frac{5}{2} \beta_0 \beta_1 r_{3,2} + 3\beta_0 r_{4,1} \nonumber\\
&& + 3\beta_0^2 r_{4,2} + \beta_0^3 r_{4,3}) \left(\frac{\alpha_{\overline{\rm MS}}(Q)}{\pi}\right)^4 +\cdots. \label{PMCexpansion}
\end{eqnarray}
where the coefficients $r_{i,0} $ for $i > 0$ are the conformal
coefficients of pQCD for $\beta=0$, and $r_{i,j } $ for $i > 0, j>0 $  are the non-conformal coefficients of the $\{\beta_i\}$-terms.

Here as for Eq.~(\ref{eq:msbar to g_1}), we have implicitly set the initial renormalization scale $\mu$ as $Q$, although as a basic property of PMC scale-setting, the determined scales of the coupling $Q_i$ at each order turn out to be minimally dependent on the initial choice of scale. Any residual initial scale dependence at finite order in pQCD is  highly suppressed, especially at the presently considered four-loop order. (One can test the initial scale dependence by recomputing the PMC predictions for  $\mu\neq Q$; this can be conveniently done by applying the RGE.)

The conformal coefficients $r_{i,0}$ are:
\begin{eqnarray}
r_{1,0} &=& \frac{3}{4}{\gamma^{\rm ns}_1}, \nonumber \\
r_{2,0} &=& \frac{3}{4}{\gamma^{\rm ns}_2}-\frac{9}{16}\big({\gamma^{\rm ns}_1}\big)^2, \nonumber \\
r_{3,0} &=& \frac{3}{4}{\gamma^{\rm ns}_3}-\frac{9}{8}{\gamma^{\rm ns}_2}{\gamma^{\rm ns}_1}+\frac{27}{64}\big({\gamma^{\rm ns}_1}\big)^3,\nonumber  \\
r_{4,0} &=& \frac{3}{4}{\gamma^{\rm ns}_4}-\frac{9}{8}{\gamma^{\rm ns}_3}{\gamma^{\rm ns}_1}-\frac{9}{16}\big({\gamma^{\rm ns}_2}\big)^2+
 \frac{81}{64} {\gamma^{\rm ns}_2} \big({\gamma^{\rm ns}_1}\big)^2-\frac{81}{256}\big({\gamma^{\rm ns}_1}\big)^4, \nonumber
\end{eqnarray}
and the non-conformal coefficients $r_{i,j}$ read:
\begin{eqnarray}
r_{2,1} &=& {3 \over 4}{\Pi^{\rm ns}_1}+{K^{\rm ns}_1},  \nonumber\\
r_{3,1} &=& {3 \over 4}{\Pi^{\rm ns}_2}+{1 \over 2}{K^{\rm ns}_2}-\frac{\gamma^{\rm ns}_1}{4}\left(\frac{3}{2}{K^{\rm ns}_1}+{9 \over 4}{\Pi^{\rm ns}_1}\right),\;\; r_{3,2}=0,  \nonumber\\
r_{4,1} &=& {3 \over 4}{\Pi^{\rm ns}_3}+{1 \over 3}{K^{\rm ns}_3}-{1 \over 4}{\gamma^{\rm ns}_1}\left({K^{\rm ns}_2}+3{\Pi^{\rm ns}_2}\right)
-\frac{\gamma^{\rm ns}_2}{4}\left({K^{\rm ns}_1}+{3 \over 2}{\Pi^{\rm ns}_1}\right) +\frac{\big({\gamma^{\rm ns}_1}\big)^2}{16} \left(3{K^{\rm ns}_1}+\frac{27}{4}{\Pi^{\rm ns}_1}\right),  \nonumber\\
r_{4,2} &=& -\frac{3}{16}\big({\Pi^{\rm ns}_1}\big)^2-{1 \over 4}{K^{\rm ns}_1} {\Pi^{\rm ns}_1}, \;\; r_{4,3} = 0, \nonumber
\end{eqnarray}
where the expressions for $\gamma^{\rm ns}_i$, $\Pi^{\rm ns}_i$ and $K^{\rm ns}_i$ are given explicitly in Refs.~\cite{Baikov:2010je, Baikov:2012zm}.

As indicated by Eq.~(\ref{PMCexpansion}), because the running of $\alpha_{\overline{\rm MS}}$ at each order has its own $\{\beta_i\}$-series as governed by the RGE, the $\beta$-pattern for the pQCD series at each order is a superposition of all of the $\{\beta_i\}$-terms which govern the evolution of the lower-order $\alpha_s$ contributions at this particular order. All known $\{\beta_i\}$-terms should be absorbed into $\alpha_{\overline{\rm MS}}$ at each order according to the RGE~\cite{Mojaza:2012mf, Brodsky:2013vpa}, thus determining its correct running behavior at each order.  Hence, after applying PMC scale-setting, only the conformal coefficients remain. The result is:
\begin{eqnarray}
\label{alpha_g1 from PMC}
\frac{\alpha_{g_1}(Q)}{\pi}=\sum_{i\geq1}{r_{i,0} \left(\frac{\alpha_{\overline{\rm MS}}(Q_i)}{\pi}\right)^i}.
\end{eqnarray}
The elimination of the divergent renormalon terms naturally leads to a pQCD series more convergent than the original one in Eq.~(\ref{PMCexpansion}). The PMC scales $Q_i$ are functions of $Q$ and read:
\begin{widetext}
\begin{eqnarray}
\ln\frac{Q_1^2}{Q^2} &=& -\frac{r_{2,1}}{r_{1,0}}-\frac{\beta_0 \left(r_{1,0} r_{3,2}-r_{2,1}^2\right)}{4r_{1,0}^2}\frac{\alpha_{\overline{\rm MS}}(Q)}{\pi}  \\
&&+\left[\frac{\beta_0^2}{16} \left(-\frac{r_{2,1}^3}{r_{1,0}^3}+\frac{2 r_{3,2} r_{2,1}}{r_{1,0}^2}-\frac{r_{4,3}}{r_{1,0}}\right) +\frac{\beta_1}{16} \left(\frac{3 r_{2,1}^2}{2 r_{1,0}^2}-\frac{3 r_{3,2}}{2 r_{1,0}}\right)\right]\left(\frac{\alpha_{\overline{\rm MS}}(Q)}{\pi}\right)^2+{\cal O}\left(\left(\frac{\alpha_{\overline{\rm MS}}}{\pi}\right)^3\right), \label{PMCscale1} \nonumber \\
\ln\frac{Q_2^2}{Q^2} &=& -\frac{r_{3,1}}{r_{2,0}}-\frac{3\beta_0\left(r_{2,0} r_{4,2}-r_{3,1}^2\right)}{8 r_{2,0}^2}\frac{\alpha_{\overline{\rm MS}}(Q)}{\pi}+{\cal O}\left(\left(\frac{\alpha_{\overline{\rm MS}}}{\pi}\right)^2\right), \label{PMCscale2} \\
\ln\frac{Q_3^2}{Q^2} &=& -\frac{r_{4,1}}{r_{3,0}}+{\cal O}\left(\frac{\alpha_{\overline{\rm MS}}}{\pi}\right) . \label{PMCscale3}
\end{eqnarray}
\end{widetext}
These expressions show that the PMC scales $Q_i$ are given as a perturbative series; any residual scale dependences in $Q_i$  is due to unknown higher-order terms. This is the first kind of residual scale dependence; the contributions from unknown high-order terms are exponentially suppressed and are thus generally small.  

A number of PMC applications have been summarized in the review~\cite{Wu:2015rga}; in each case the PMC works successfully and leads to improved agreement with experiment.  Furthermore, this multi-scale PMC approach corresponds to the fact  that separate renormalization scales and effective numbers of quark flavors appear for each skeleton graph. The coefficients of the resulting pQCD series match the coefficients of the corresponding conformal theory with $\beta=0$, ensuring the scheme-independence of the PMC predictions at any fixed order.

For convenience, we provide the conformal coefficients $r_{i,0}$ and PMC scales $Q_i$ after substitution of the $\gamma^{\rm ns}_i$, $\Pi^{\rm ns}_i$ and $K^{\rm ns}_i$ into Eq.~(\ref{PMCexpansion}). They are, up to four-loop order:
\begin{widetext}
\begin{eqnarray}
r_{1,0} &=& 1, \nonumber  \\
r_{2,0} &=& 1.6042-0.1528n_f,  \nonumber  \\
r_{3,0} &=& 5.5335-1.7370n_f-0.01980n_f^2, \nonumber  \\
r_{4,0} &=& 21.5613-8.4884n_f+0.5050n_f^2   +0.004503n_f^3 \nonumber,
\end{eqnarray}
\begin{eqnarray}
\ln\frac{Q_1^2}{Q^2} &=& -1.08333+(3.2274-0.1956n_f)\frac{\alpha_{\overline{\rm MS}}(Q)}{\pi} \nonumber \\
&&+(1.6076-0.2282n_f-0.03532n_f^2)\left(\frac{\alpha_{\overline{\rm MS}} (Q)}{\pi}\right)^2 +{\cal O}\left(\left(\frac{\alpha_{\overline{\rm MS}}}{\pi}\right)^3\right), \nonumber \\
\ln\frac{Q_2^2}{Q^2} &=& -\frac{5.2728-0.5918n_f}{1.6042-0.1528n_f}-\frac{0.08756 n_f^3-3.00195 n_f^2+32.6111 n_f-114.146}{0.02334 n_f^2-0.4902 n_f+2.5734}\frac{\alpha_{\overline{\rm MS}}(Q)}{\pi} \nonumber \\
&& +{\cal O}\left(\left(\frac{\alpha_{\overline{\rm MS}}}{\pi}\right)^2\right), \nonumber \\
\ln\frac{Q_3^2}{Q^2} &=& -\frac{44.1983-8.7216 n_f+0.2165 n_f^2}{5.5335-1.7371 n_f-0.01980 n_f^2}+{\cal O}\left(\frac{\alpha_{\overline{\rm MS}}}{\pi}\right) \nonumber.
\end{eqnarray}
\end{widetext}
The PMC scale of the last known order, $Q_4$, remains undetermined because
the five-loop and higher order $\{\beta_i\}$-terms are unknown. As a test, we can
set $Q_4=Q_3$ or $Q_4=Q$, which leads to the second kind of residual scale
dependence. This scale dependence, however, generates negligible uncertainty. For example, we have computed
$\alpha_{g_1}(Q)$ using both prescriptions, and the results are nearly identical
because of the fast convergence of the PMC series.

We note that the small values of $Q$ (around 1 GeV), with $n_f=3$  lead to an almost zero $Q_3$; this reflects the fact that in
the soft $Q$-region, the intermediate gluons are effectively
nonperturbative, and thus information on  the behavior of $\alpha_s$  at low momentum is required.

We shall adopt a natural extension of the perturbative $\alpha_s$-running
behavior as determined from  the high $Q$-region.   Then, to avoid having $Q_3$ enter the nonperturbative
region,  we will use as the alternative scale $Q_3 = 40 \times Q$~\cite{pmc-scale}.
Although  we have also performed the calculations for values of $n_f$ determined by the
PMC scale $Q_i$, we will use $n_f=3$ for the results in the next sections in order to
compare meaningfully with the results reported in Refs.~\cite{Deur:2014qfa, Deur:2016cxb, Deur:2016opc}.

The results in this article use $\alpha_{g_1}(Q)$ computed
with the scales $Q_i$ calculated up to next-to-next leading order.
However, for reference, we  also provide here their values for $n_f = 3$ and at leading order:
\begin{eqnarray}
Q_1 &=& 0.581 Q, \nonumber  \\
Q_2 &=& 0.217 Q, \nonumber  \\
Q_3 &=& 40 \, Q. \nonumber
\end{eqnarray}

It is informative to compare the coefficients $a_i$ obtained from the conventional pQCD series,
Eq.~(\ref{eq:msbar to g_1}), to  the PMC coefficients $r_{i,0}$.  The $a_i$ values for $n_f=3$
are \cite{Baikov:2010je, Kataev:2005hv}:
\begin{eqnarray}
a_{1} &=& 1,  \nonumber \\
a_{2} &=& 3.58,  \nonumber \\
a_{3} &=& 20.21, \nonumber  \\
a_{4} &=& 175.7, \nonumber  \\
a_{5} &\sim & 893.38, \label{convseries}
\end{eqnarray}
which can be compared with the $r_{i,0}$ for $n_f=3$:
\begin{eqnarray}
r_{1,0} &=& 1,  \nonumber\\
r_{2,0} &=& 1.14583,  \nonumber\\
r_{3,0} &=& 0.144097,  \nonumber\\
r_{4,0} &=& 0.762723,  \label{PMCseries}
\end{eqnarray}
The $a_i$ values become very large at high orders, a manifestation of the factorial renormalon
growth $({\alpha_s / \pi})^n \beta_0^n n!$ of pQCD series using conventional scale setting. In
contrast, the conformal coefficients $r_{i,0}$ have reasonable values of  order $1$, as expected from the PMC procedure. This much-improved convergence allows for more precise predictions.

\section{Comparisons  of the PMC and  conventional predictions for the Bjorken sum rule\label{PMC test}}

The PMC approach can be tested by comparing $\alpha_{g_1}(Q)$ computed using the PMC prediction (\ref{alpha_g1 from PMC}) versus the conventional pQCD calculation (\ref{eq:msbar to g_1}). In each case, the prediction will be estimated up to fourth order and with $n_f=3$.  For these computations, we will evaluate $\alpha_{\overline{\rm MS}}$ up to five loops assuming $\Lambda_{\overline{\rm MS}}^{(n_f=3)} = 0.332(17)$ GeV~\cite{Olive:2016xmw}, which is the current  world average from various experimental and lattice QCD data using $\chi^2$ minimization.

\begin{figure}[tb]
\includegraphics[width=8.6cm]{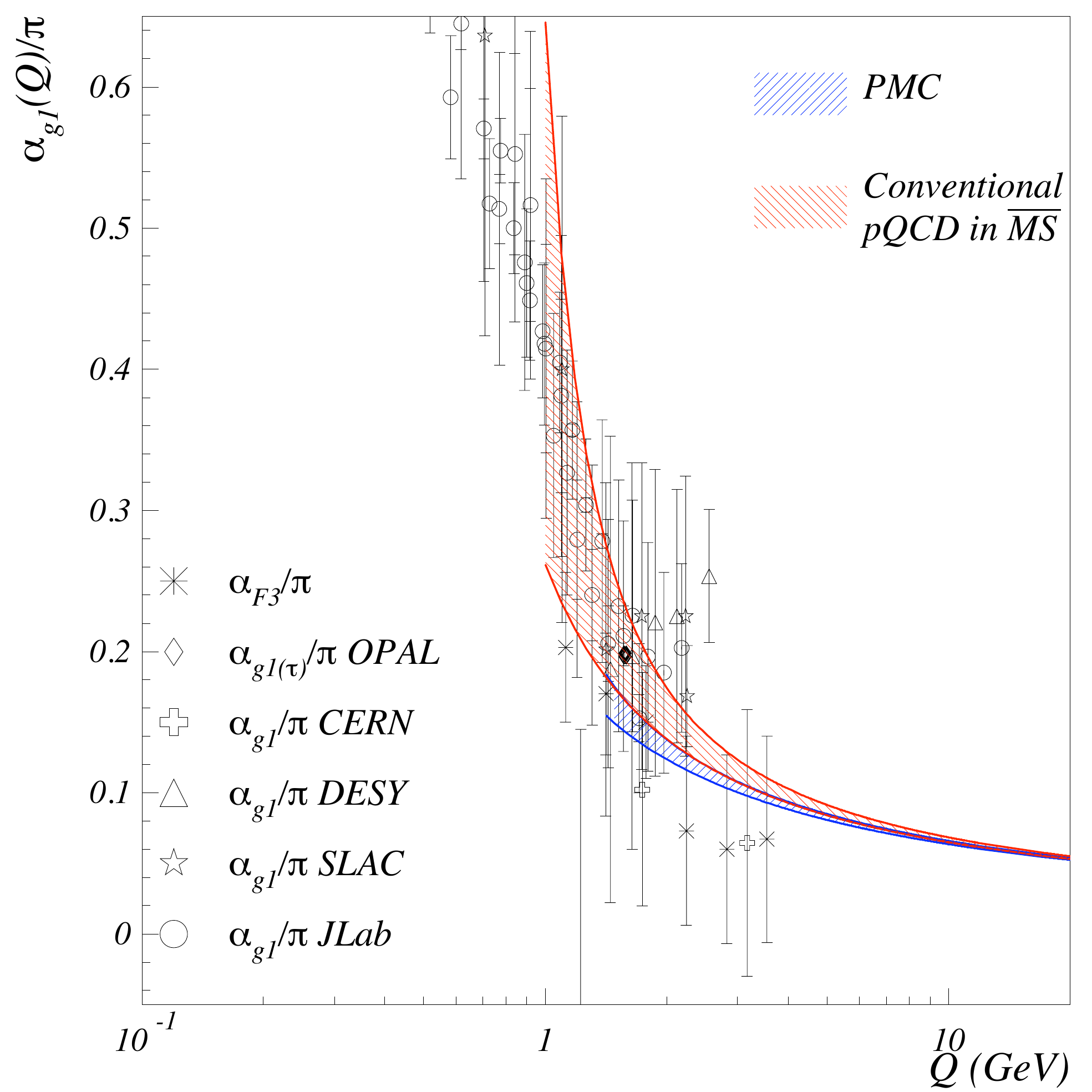}
\caption{(Color online) the PMC and conventional $\overline{\rm MS}$ predictions for $\alpha_{g_1}(Q)/\pi$ computed for $n_f=3$, with $\Lambda_{s,\overline{\rm MS}}^{(n_f=3)} = 0.332(17)$~\cite{Olive:2016xmw}. The various symbols represent the experimental determinations of $\alpha_{g_1}(Q^2)$ or $\alpha_{F_3}(Q^2)$. }
\label{Fig:PMC_BjSR_comp}
\end{figure}

In Fig.~\ref{Fig:PMC_BjSR_comp}, we display $\alpha_{g_1}(Q)/\pi$ calculated using the RS-independent PMC prediction versus the conventional pQCD in the $\overline{\rm MS}$-scheme, together with the available experimental data~\cite{Deur:2005cf}.  We also show the experimental data for $\alpha_{F_3}(Q)$, since the two effective charges $\alpha_{F_3}$ and $\alpha_{g_1}$ are in practice nearly identical~\cite{Deur:2005cf}.   We compute $\alpha_{g_1}(Q)$ for values of the argument of $\alpha_{\overline{\rm MS}}(\mu)$ greater than 1 GeV, $\mu>1$ GeV.  In the conventional pQCD prediction of $\alpha_{g_1}(Q)$ %in the $\overline{\rm MS}$-scheme,
the renormalization scale is directly set to $Q$ and $\alpha_{g_1}(Q)$ is computed for $Q > 1$ GeV.  For the PMC scale-setting  calculation, $\mu>1$ GeV implies that $\alpha_{g_1}(Q)$ is computed for $Q > 1.48$ GeV, the reason for which will be discussed in the next subsection.

The total uncertainties of the two predictions stem from several sources:
\begin{itemize}

\item The uncertainty of the perturbative approximant for $\alpha_{\overline{\rm MS}}$, which we estimate by taking the difference between the expressions of $\alpha_{\overline{\rm MS}}$ at order $\beta_3$ and at order $\beta_4$.

\item The 17 MeV uncertainty on the value of  $\Lambda_{\overline{\rm MS}}^{(n_f=3)}$~\cite{Olive:2016xmw};

\item The truncation uncertainty in the PMC series (\ref{alpha_g1 from PMC}) or in the conventional $\overline{\rm MS}$ series (\ref{eq:msbar to g_1}). For the PMC series, it is estimated by taking the difference between the fourth order and third order terms: $\big(\alpha_{\overline{\rm MS}}/\pi\big)^3 \big(r_{4,0} \alpha_{\overline{\rm MS}}/\pi - r_{3,0}\big)$. For the conventional $\overline{\rm MS}$ pQCD series, it is taken as the difference between the estimated fifth order term and the calculated fourth order term: $\big(\alpha_{\overline{\rm MS}}/\pi\big)^4 \big(a_{5} \alpha_{\overline{\rm MS}}/\pi - a_{4}\big)$.
\end{itemize}

Fig.~\ref{Fig:PMC_BjSR_comp} shows that the four-loop PMC and the conventional
pQCD calculations of $\alpha_{g_1}(Q)$ are consistent with each other, although only
marginally for $Q$ below a few GeV.  

We have also performed the same  calculations
by  computing the value of the quark flavor variable $n_f$,  according to the quark mass threshold as determined by the value of $Q$,
in the case of the conventional pQCD calculation of $\alpha_{g_1}(Q)$, or the values of the $Q_i$ PMC scales for the PMC calculation. The results are similar to that shown in Fig.~\ref{Fig:PMC_BjSR_comp}.

A notable feature in Fig.~\ref{Fig:PMC_BjSR_comp} is that the theoretical uncertainty of the 
PMC prediction is significantly smaller than that of the conventional pQCD prediction. As 
seen from Eqs.~(\ref{convseries}) and (\ref{PMCseries}), this is due to the fact 
that the pQCD series using PMC scale-setting converges much faster than the conventional pQCD series.

\section{Matching to the nonperturbative domain\label{matching}}

In Refs.~\cite{Deur:2014qfa, Deur:2016cxb}, a method has been proposed to relate the perturbative QCD asymptotic scale
$\Lambda_s$  to the  hadron mass  scale such as the proton mass.   The scale $Q_0$ which signifies the transition
between the perturbative and  nonperturbative domains of QCD is also 
determined by
this method. Both $\Lambda_s$ and $Q_0$ are obtained  
in any renormalization scheme in the pQCD domain.
This method uses  the analytic form of $\alpha_{g_{1}}$  \cite{Brodsky:2010ur}  predicted in the nonperturbative domain
by Light Front Holographic QCD (LFHQCD)~\cite{Brodsky:2014yha}:
\begin{equation} \label{alphaIR}
{ \alpha_{g_{1}}\left(Q \right) }  = \pi \exp \left( - {Q^2 \over 4\kappa^2} \right),
\end{equation}
where $\kappa$ is a  universal nonperturbative scale derived  from
hadron masses, for example, $\kappa = M_{\rho}/\sqrt{2} =0.548$ GeV, where $M_{\rho}$ is the mass of the $\rho$--meson.
Alternatively, $\kappa$ can be obtained from fits to hadron form-factors, the Regge slopes, or the Bjorken sum rule Eq.~(\ref{eqn:bj}).
Although the value of $\kappa$ is universal, in practice, the approximations used in LFHQCD induce a $\simeq 10\%$ variation. 
The latest determination gives $\kappa = 0.523(24)$ GeV~\cite{Brodsky:2016yod}. The Gaussian form Eq.~(\ref{alphaIR}) is in excellent agreement with data and the various
nonperturbative calculations of $\alpha_s(Q)$~\cite{Deur:2016cxb,Deur:2016tte}, including the recent result based on  
Schwinger-Dyson Equations~\cite{Binosi:2016nme}.

The basis for the matching procedure to determine $\Lambda_s$ is the overlap of the
domains of applicability of LFHQCD ($Q \lesssim 1.3$ GeV) with the pQCD
($Q \gtrsim 1.0$ GeV)~\cite{LFHQCD_applicability}.  Continuity of $\alpha_{g_{1}}(Q)$  and its first derivative
implies that Eq.~(\ref{eq:msbar to g_1}) and Eq.~(\ref{alphaIR}), as well as their
corresponding $\beta$--functions, can be equated in the overlap region.  The simultaneous solution to these  two equations  provides an analytical relation between $\Lambda_s$ and $\kappa$, as well as the transition scale $Q_0$. This leads to a determination of $\Lambda_{\overline{\rm MS}}^{(n_f=3)}=0.339(19)$ GeV~\cite{Deur:2016opc}, with a precision {\it on par} with that of the averaged world data of 0.332(17) GeV~\cite{Olive:2016xmw}.

Since the PMC provides a more precise determination of
$\alpha_{g_{1}}(Q)$ than conventional renormalization scale-setting, it is interesting to investigate if the procedure is also applicable
using Eq.~(\ref{alpha_g1 from PMC}) rather than Eq.~(\ref{eq:msbar to g_1}) to
improve the determination of $\Lambda_s$.

\begin{figure}[htb]
\includegraphics[width=8.6cm]{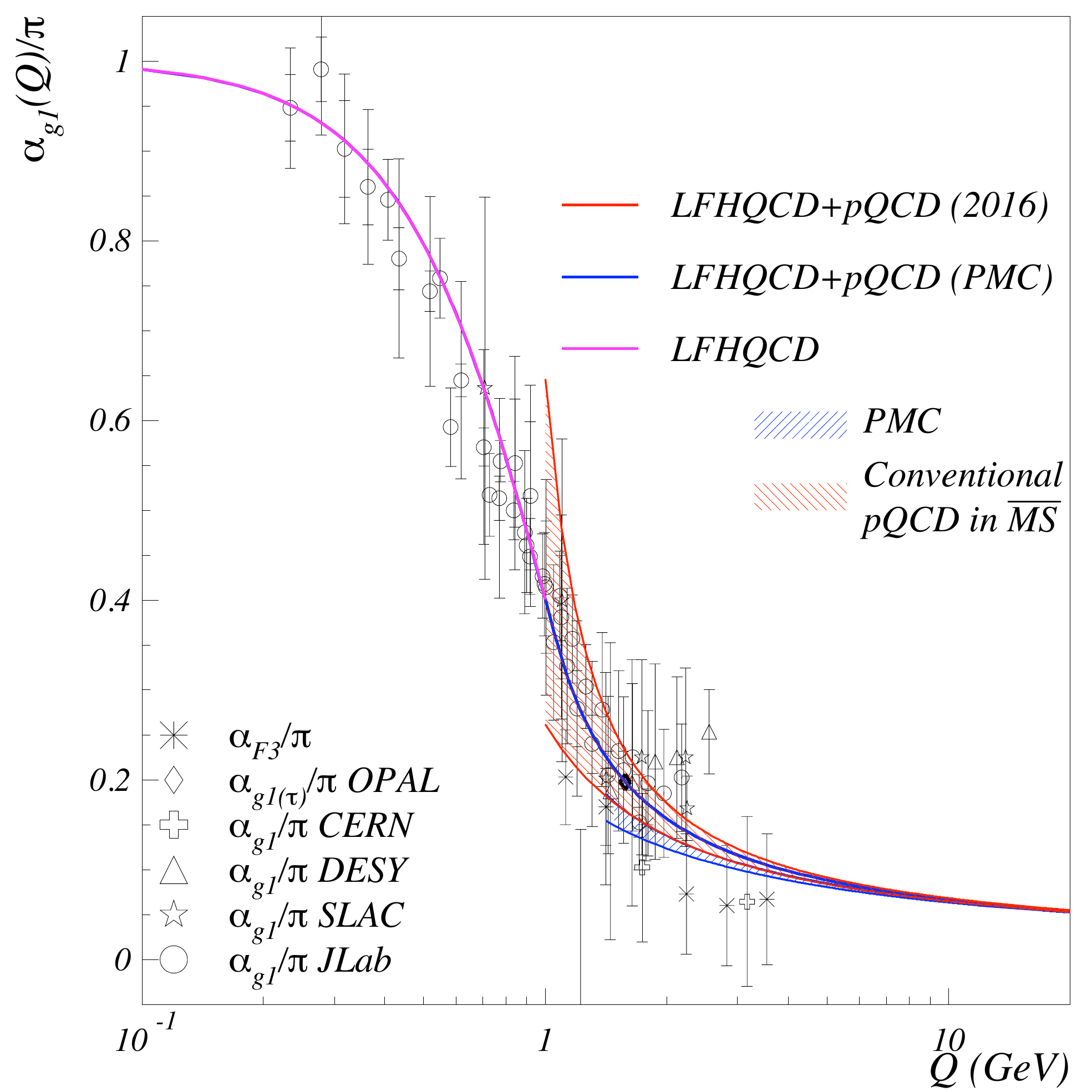}
\caption{(Color online) Matching procedure applied to PMC calculation (blue line). It is matched to the LFHQCD results (magenta line) by requiring the continuity of both $\alpha_{g_{1}}$ and its $\beta$-function. The blue band is the PMC prediction evaluated down to $Q=1.5$ GeV, and the red band shows the conventional $\overline{\rm MS}$ prediction from Ref.~\cite{Deur:2016opc}. We use $\kappa = 0.523$ GeV for LFHQCD.}   \label{Fig:matching}
\end{figure}

Following the same matching procedure, we 
 have computed the PMC prediction using $\kappa = 0.523$ GeV.
To reach the matching point $Q_0$, it is necessary to extrapolate the PMC prediction down
to $Q=1$ GeV, which implies that the $\alpha_{\overline{\rm MS}}(\mu)$ must be extrapolated
down to $\mu = 0.68$ GeV. The result is shown in Fig.~\ref{Fig:matching}.
As a comparison, we also show the conventional $\overline{\rm MS}$ prediction~\cite{Deur:2016opc} in the figure.

The matching of the PMC prediction to LFHQCD yields a large value for
$\Lambda_{s,\overline{\rm MS}}^{(n_f=3)}$ = 0.406(17) GeV. This explains why,
compared to Fig.~\ref{Fig:PMC_BjSR_comp}, a better agreement between the
matched PMC curve (blue line) and the conventional $\overline{\rm MS}$ pQCD
calculations (red band) is observed in Fig.~\ref{Fig:matching}.

The determined transition scale, $Q_0=1.14$ GeV, is below the scale at which the present PMC calculation is applicable ($Q \approx 1.48$ GeV). The failure of this self-consistency check indicates that the matching procedure cannot be used with the PMC calculation, at least when $\overline{\rm MS}$ is used as an auxiliary RS. This explains why the matching procedure yields $\Lambda_{\overline{\rm MS}}^{(n_f=3)}=0.406(17)$ GeV, which is somewhat larger than the world data. This is reflected in Fig.~\ref{Fig:matching} by the fact that the blue line does not lie within the blue band.

In the case of conventional scale-setting, the renormalization scale $\mu$ is
fixed at its initial value $Q$.   In contrast, as shown by
Eqs.~(\ref{PMCscale1}) and (\ref{PMCscale3}), the determined PMC
scale $Q_i$ for each order is a function of $Q$ which can result in $Q_i$ scales that are
larger or smaller than $Q$. This has consequences for the matching procedure
proposed in Ref.~\cite{Deur:2014qfa}{\color{blue},}  which requires that  the transition
between nonperturbative and perturbative QCD
occurs at a point $Q_0$ rather than over a non-zero $Q$ range.

In the case of conventional scale-setting, the meaning of the inflection point  $Q_0$ is unambiguous: $\alpha_s(Q)$ has  perturbative behavior for $Q>Q_0$ and  nonperturbative
behavior for $Q< Q_0$. These are determined by pQCD and LFHQCD, respectively.
On the other hand, in the case of the PMC scale-setting, some  PMC scales are smaller
than the determined $Q_0$, thus leading to an apparent incompatibility; i.e.,
if the determined PMC scale $Q_i$ is  less than $Q_0$, the meaning of $Q_0$ is questionable
since $Q_i$ is now within the nonperturbative region. This is indeed the case for the present
procedure. Thus{\color{blue},}  due to the fast convergence of PMC series, we have
$\alpha_{g_1}(Q) \sim \alpha_{\overline{\rm MS}}(Q_1)$, where the PMC scale $Q_1 = 0.45$ GeV
is significantly smaller than the transition scale $Q_0=1.14$ GeV. This conflict  could be due to the fact that some of nonperturbative effects which are not accounted for in the (perturbative) derivation of the PMC scales $Q_i$, such as those from the high-twist terms~\cite{high-twist}, may have already come into the higher-order calculations. For example, the renormalization scale for the heavy-quark loop which appears in the three-gluon coupling depends nontrivially on the virtualities of the three gluons entering the three-gluon vertex~\cite{Binger:2006sj}.

\begin{figure}[htb]
\includegraphics[width=8.6cm]{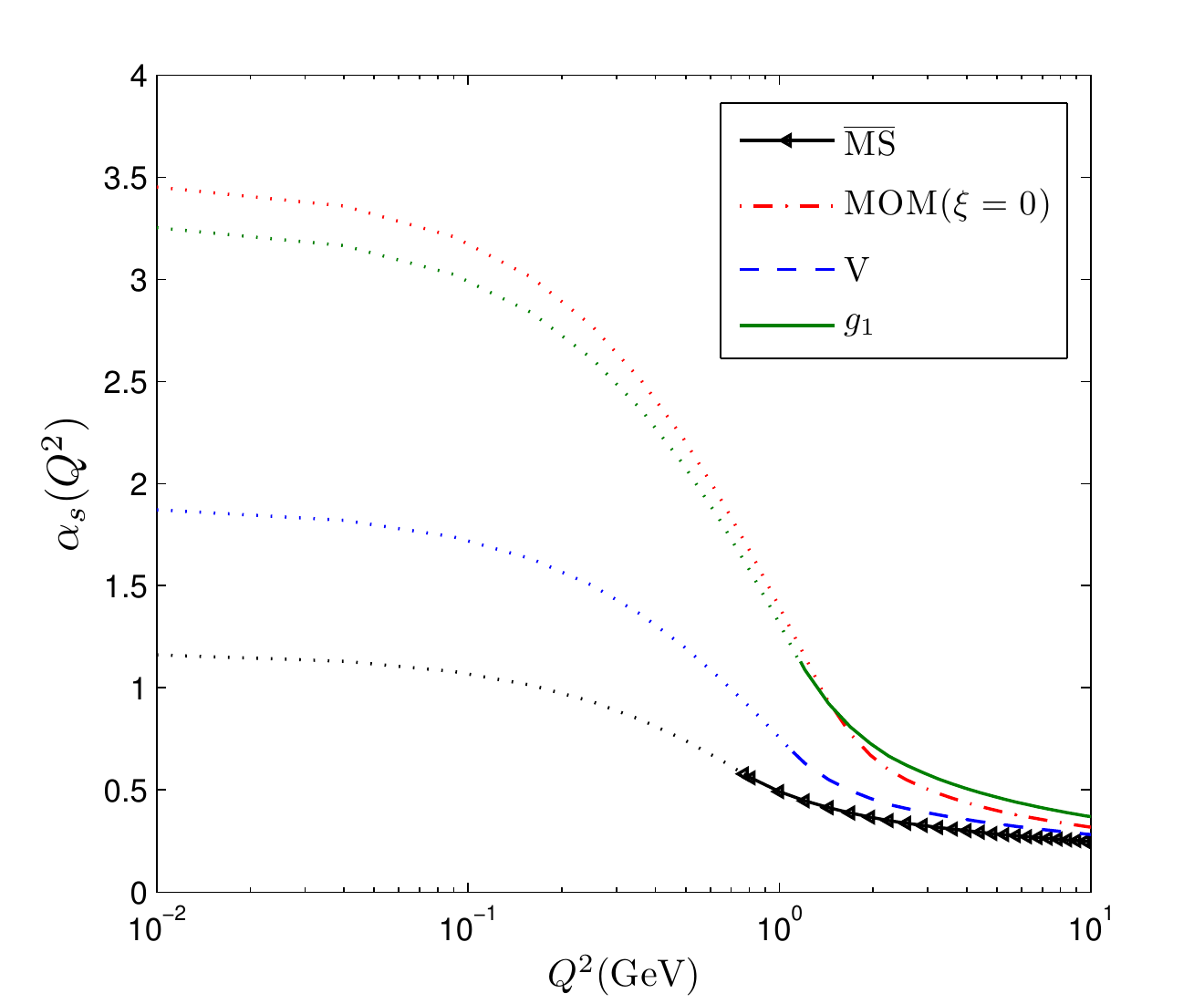}
\caption{\label{Fig:alphas} (Color online) The strong coupling $\alpha_s(Q^2)$ for various renormalization schemes~\cite{Deur:2016cxb}. The lines in the perturbative region are the perturbative calculations done at
order $\beta_3$. The dashed curves are their matched LFHQCD continuations into the nonperturbative domain.}
\label{alphas-running}
\end{figure}

This problem may be solved by transforming to a different ${\rm MS}$-like
scheme;  e.g.,  the $R_\delta$ scheme~\cite{Mojaza:2012mf, Brodsky:2013vpa} 
where the subtraction $\ln 4\pi -\Gamma_E -\delta$ is used within the minimal subtraction procedure.
(The conventional $\overline{\rm MS}$-scheme is the $R_\delta$-scheme
corresponding to $\delta=0$.) The scheme transformation between different
$R_\delta$-schemes corresponds simply to a displacement of their
corresponding scales; $\mu_\delta^2 = \mu_{\overline{\rm
MS}}^2\  \exp({\delta})$; thus a proper choice of $\delta$ may avoid the
small scale problem found for the $\overline{\rm MS}$-scheme. This problem may
also be solved by using a different auxiliary RS,
such as the MOM scheme with $\xi=0$ (Landau gauge)~\cite{Celmaster:1979km},
or the $V$ scheme~\cite{Peter:1996ig}. This possibility is motivated by comparing the running
behaviors of $\alpha_s$  for  different  schemes; examples are presented
in Fig.~\ref{alphas-running}. It shows that to ensure the scheme-independence of the couplings, e.g. $\alpha_{\overline{\rm MS}}(\mu_{\overline{\rm MS}})=\alpha_{\rm V}(\mu_V)=\alpha_{\rm MOM}(\mu_{\rm MOM})$, we must have $\mu_{\overline{\rm MS}}<\mu_{\rm V}<\mu_{\rm MOM}$. This fact has been observed by the LO commensurate scale relations among different effective couplings~\cite{Brodsky:1994eh}. Thus a larger PMC scale can be achieved when the $V$-scheme or MOM-scheme is adopted as the auxiliary RS. For example, in  the  case of the $V$-scheme, the PMC prediction is applicable down to $Q=1$ GeV if the perturbative behavior of $\alpha_{\overline{\rm MS}}(\mu)$ can be extrapolated down to $\mu = 0.85$ GeV~\cite{Brodsky:1994eh}, which is larger than the corresponding value of $\mu = 0.68$ GeV required for the $\overline{\rm MS}$-scheme.

Another avenue to address the problem could be to use the single-scale approach for the PMC~\cite{Shen:2017pdu}, where a single effective scale replaces the individual PMC scales in the sense of a mean value theorem; this can avoid the small scale problem which can appear at specific orders in the multi-scale PMC approach. These investigations will be reported in a future publication.

\section{Summary and conclusion}

In this paper, we have tested the PMC scale-setting procedure by comparing its prediction in the $\overline{\rm MS}$ scheme
for  the effective charge $\alpha_{g_{1}}(Q)$ defined from the Bjorken sum rule with the prediction obtained using conventional renormalization scale-setting. To this
end, we have calculated the necessary PMC coefficients and renormalization scales. We have verified that the PMC series converges
much faster than the conventional $\overline{\rm MS}$ pQCD series, which results in a significantly
smaller uncertainty for the PMC pQCD prediction. Thus the central objective of the PMC is
realized: it provides a determination of $\alpha_{g_{1}}(Q)$ compatible with the data and the
conventional pQCD calculation, but without scheme-dependence and with significantly improved precision.

 As an important application, we have investigated the possibility of determining
 $\Lambda_{\overline{\rm MS}}$ from hadronic scales by matching the PMC calculation for pQCD  to the nonperturbative light-front holographic QCD prediction for $\alpha_{g_{1}}(Q)$.
This had been done previously using the conventional scale-setting pQCD prediction; this worked well, giving $\Lambda_{\overline{\rm MS}}^{(n_f=3)} = 0.339(19)$ GeV.
However, we have found that the domain of  applicability of the nonperturbative LFHQCD and the domain of  applicability of the perturbative PMC predictions do not overlap if the 
$\overline{\rm MS}$-scheme is used as the auxiliary scheme, causing the matching procedure to fail.  This problem arises from the fact that the PMC scales at certain orders in the 
$\overline{\rm MS}$-scheme are{\color{blue},} in some cases{\color{blue},} smaller than the  transition scale $Q_0$.  A detailed investigation for solving this problem,  
using alternative renormalization schemes and/or the single-scale PMC procedure for pQCD, is in preparation.

\acknowledgments{This paper is based upon work supported by the U.S. Department of Energy, the Office of Science, and the Office of Nuclear Physics under contract DE-AC05-06OR23177. This work is also supported by the Department of Energy contract DE--AC02--76SF00515 and by the National Natural Science Foundation of China under Grant No.11625520.  SLAC-PUB-16959, JLAB-PHY-17-2394}.


\begin{thebibliography}{00}
\bibitem{pmc-scale}If the $Q$ value ensures that $n_f=4$, e.g. $Q \geq m_c\sim 1.275 \, {\rm GeV}$,
we get a reasonable $Q_3 \simeq 40 \, Q$.
\bibitem{LFHQCD_applicability} The value of 1.3 GeV is determined as the $Q$ value where the LFHQCD prediction for $\alpha_{g_1}$ starts to disagree by more than 10\% with the central value of $\alpha_{g_1}$ obtained using conventional pQCD in the $\overline{\rm MS}$ RS, up to 4 loops for the $\beta$-series and 4th order in the Bjorken sum. The 10\% prescription is chosen as typical of the general uncertainty on $\alpha_{g_1}$. Likewise, the value of 1.0 GeV for the lower limit of applicability of conventional pQCD in the $\overline{\rm MS}$ RS  is determined as the value where  $\alpha_{g_1}$ from conventional pQCD is 10\% larger than the LFHQCD prediction. This agrees with the typical prescription that pQCD is applicable for $Q>1$ GeV.
\bibitem{high-twist} At high-orders some of the propagators which share the typical momentum flow of the process could be soft, leading to nonperturbative high-twist contributions;
\end{thebibliography}
\end{document}